\documentclass[fleqn,usenatbib]{mnras}
\usepackage{epsfig,amsmath}
\usepackage{color,subfigure}

\usepackage{mathrsfs,amssymb,amstext,amssymb}
\usepackage{ulem}
\usepackage{url}
\usepackage{bm}
\usepackage{newtxtext,newtxmath}

\usepackage[T1]{fontenc}
\usepackage{ae,aecompl}

\usepackage{graphicx}	
\usepackage{subfigure}
\usepackage{booktabs}

\title[Galaxy properties in the cosmic web of EAGLE simulation]{Galaxy properties in the cosmic web of EAGLE simulation}

\author[Wenxiao Xu et al.]{
Wenxiao Xu,$^{1,2}$
Qi Guo,$^{1,2}$\thanks{E-mail: guoqi@nao.cas.cn}
Haonan Zheng,$^{1,2}$
Liang Gao,$^{1,2}$\thanks{E-mail: lgao@bao.ac.cn}
Cedric Lacey,$^{3}$
Qing Gu,$^{1,2}$\newauthor
Shihong Liao,$^{1}$
Shi Shao,$^{3}$
Tianxiang Mao,$^{1,2}$
Tianchi Zhang$^{1,2}$
and Xuelei Chen$^{1,2}$
\\
$^{1}$Key Laboratory for Computational Astrophysics, National Astronomical Observatories, Chinese Academy of Sciences, Beijing, 100012, China\\
$^{2}$University of Chinese Academy of Sciences, Beijing, 100049, China\\
$^{3}$Institute for Computational Cosmology, Department of Physics, Durham University, South Road, Durham, DH1 3LE, UK\\
}

\date{Accepted XXX. Received YYY; in original form ZZZ}

\pubyear{2020}

\begin{document}
\label{firstpage}
\maketitle

\begin{abstract}
We investigate the dependence of the galaxy properties on cosmic web environments using the most up-to-date hydrodynamic simulation:  Evolution and Assembly of Galaxies and their Environments (EAGLE). The baryon fractions in haloes and the amplitudes of the galaxy luminosity function decrease going from knots to filaments to sheets to voids. Interestingly, the value of L$^*$ varies dramatically in different cosmic web environments.  At z = 0, we find a characteristic halo mass of $10^{12} h^{-1}\rm M_{\odot}$, below which the stellar-to-halo mass ratio is higher in knots while above which it reverses. This particular halo mass corresponds to a  characteristic stellar mass of $1.8\times 10^{10} h^{-1}\rm M_{\odot}$. Below the characteristic stellar mass central galaxies have redder colors, lower sSFRs and higher metallicities in knots than those in filaments, sheets and voids, while above this characteristic stellar mass, the cosmic web environmental dependences either reverse or vanish. Such dependences can be attributed to the fact that the active galaxy fraction decreases along voids, sheets, filaments and knots. The cosmic web dependences get weaker towards higher redshifts for most of the explored galaxy properties and scaling relations, except for the gas metallicity vs. stellar mass relation. 
\end{abstract}




\section{Introduction}
The large scale structure (LSS) of the Universe exhibits a web-like structure, which is usually categorized into four components: voids, sheets, filaments and knots. It originates from small perturbations in the very early Universe and is shaped gradually by the large scale gravitational fields. Different components correspond to different stages of gravitational collapse. Large sheets of matter form via gravitational collapse along one principal direction, filaments form via gravitational collapse along two principal axes and knots form via gravitational collapse along three principal axes. 
The relatively empty regions of the Universe between knots, filaments and sheets are referred to as cosmic voids, whose density is well below the cosmic mean value. Most mass is contained in knots and filaments, while most volume is filled by voids. The relative mass fractions and filling factors of each component could evolve significantly from high to low redshifts  \citep{2019Cui, 2017ApJ...838...21Z}.

The cosmic web environments have influences on the halo formation history and the halo properties. Haloes in knots tend to be older than those in other web environments, and massive haloes in knots have higher spin than those in filaments, sheets and voids \citep[e.g.][]{Hahn2007}. The halo mass function strongly depends on the web environment such that it has a much higher amplitude in filaments than in voids \citep[e.g.][]{2018Ganeshaiah}. Recently, the large-scale cosmic web environment was also found to correlate with the orientation of haloes \citep[e.g.][]{Zhang_2009,2012Codis,2018Ganeshaiah}.
However, the role the cosmic web plays in galaxy formation is not clear. Recent work found that the blue galaxy fraction and the star formation rate of blue galaxies decline from the field to knots at low redshift, but the dependence disappears at higher redshifts \citep{Darvish_2017,2018Kraljic}. Similar trends are also found in the state-of-the-art hydro-dynamical simulation HORIZON-AGN  \citep{Dubois:2014lxa}.
Using the Sloan Digital Survey (SDSS), \citet{2017MNRAS.466.1880C} demonstrated that red, massive galaxies tend to reside in filaments more than blue, less-massive galaxies.   
 
It has been suggested that the cosmic web shapes the galaxy properties primarily through the strong dependence of the galaxy properties on the mass of their host halo  \citep[e.g.][] {2010MNRAS.404.1801G, Goh2018, Yan2013}. 
This is also supported by   \citet{Eardley2015} who found in Galaxy And Mass Assembly survey (GAMA) that the strong variation of galaxy properties in different cosmic web structures vanishes when comparing at the same local density. 
However, by using SDSS data, \citet{Poudel2017} found that at a fixed halo mass, central galaxies in filaments have redder colors, higher stellar mass, lower specific star formation rates and higher abundances of elliptical galaxies compare to those outside the filaments. Observationally it is hard to measure the halo mass for an individual galaxy while stacking methods could smear out the cosmic web dependence if it is not strong enough. Theoretical work using cosmological hydro-dynamical simulations can provide more clues to such studies.  Using IllustrisTNG, \citet{2019Martizzi}  found at a given halo mass, galaxies with stellar masses lower than the median value are more likely to be found in voids and sheets, whereas galaxies with stellar masses higher than the median are more likely to be found in filaments and knots. \citet{Liao2018} found that haloes in filament have higher baryon fractions and stellar mass fractions compare to those in the field.

In recent years, cosmological hydro-dynamical simulations have gained great success in reproducing many observed galaxy properties \citep[e.g.][]{Vogelsberger2014, Schaye2015}. In such simulations the cosmic web effect is taken into account automatically. Here we use the state-of-art cosmological hydrodynamical simulation, Evolution and Assembly of Galaxies and their Environments (EAGLE) to disentangle the connections between galaxies, dark matter haloes and the cosmic web and revisit the relation between galaxy properties and geometric cosmic web structures. 
  
This paper is organized as follows.  We introduce the EAGLE simulation and describe the environmental classification method in Section~\ref{sec:simulation}. The cosmic-web dependence of the various baryonic components and the scaling relations are presented in Section~\ref{sec:baryon} and Section~\ref{sec:scaling}, respectively. In Section~\ref{sec:conclusions}, we summarize our results.

\section{SIMULATION AND METHODS}
\label{sec:simulation}

\begin{figure*}
\centering
\begin{minipage}[!htbp]{0.45\linewidth}
\includegraphics[width=3.2in]{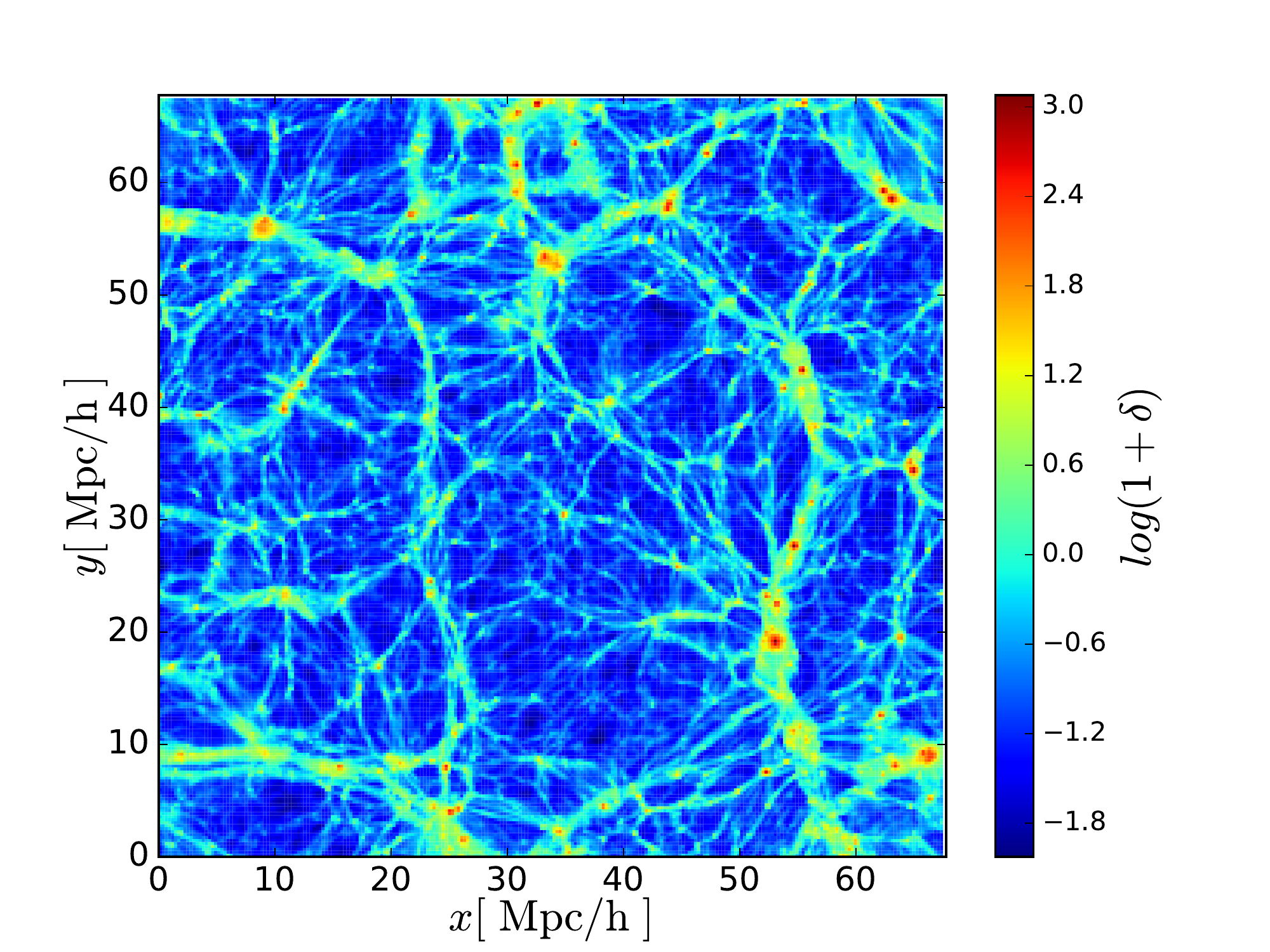}
\end{minipage}%
\begin{minipage}[!htbp]{0.45\linewidth}
\includegraphics[width=3.2in]{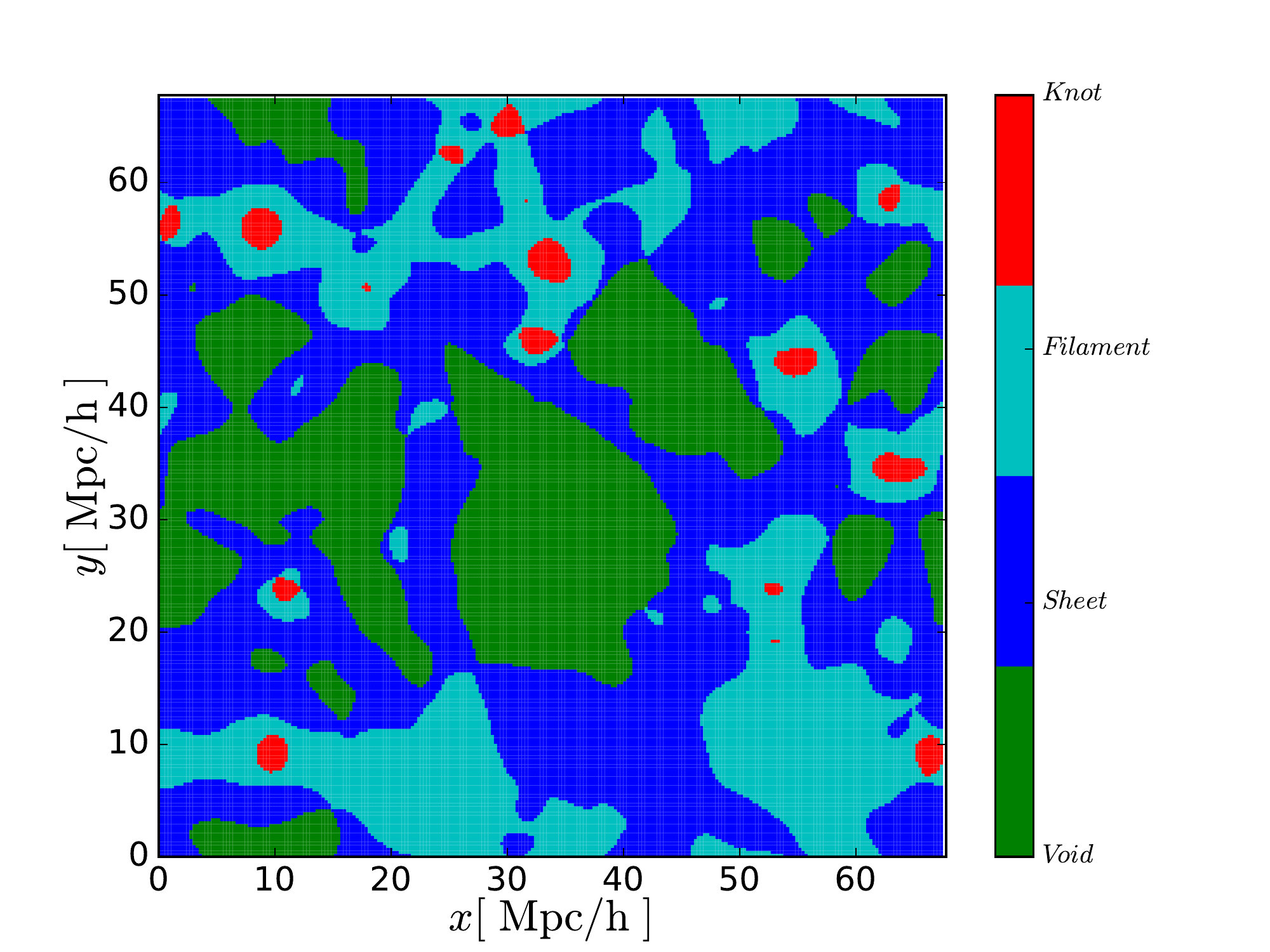}
\end{minipage}%
\caption{Left panel: Overdensity map of a slice of $0.265\rm Mpc/h$ thickness. The color bar denotes the values of logarithm of the overdensity. Right panel:  The corresponding cosmic web elements generated using the method as described in Section \ref{sec:environment}. Red, cyan, blue and green regions are for the knots, filaments, sheets and voids, respectively.}
\label{fig:web}
\end{figure*}

\subsection{EAGLE simulation}
The EAGLE simulation consists of a series of cosmological simulations performed with a modified version of the N-body Tree-PM smoothed particle hydrodynamics (SPH) code GADGET-3 \citep{Springel2005}. In this paper we use the largest volume EAGLE simulation, labelled as L100N1504, which was carried out in a box of 100Mpc each side, tracing $1504^3$ dark matter particles and an equal number of baryonic particles. The initial mass of the gas particles is $1.81 \times 10^6 \rm M_{\odot}$ and the dark matter particles mass is $9.70 \times 10^6 \rm M_{\odot}$. The EAGLE simulations adopts the cosmological parameters taken from the Planck results \citep{Planck2014}: $\rm \Omega_m = 0.307$, $\rm \Omega_{\Lambda} = 0.693$, $\rm \Omega_b = 0.04825$, $h = 0.6777$, $\sigma_8 = 0.8288$, $n_s = 0.9611$ and $\rm Y = 0.248$.  

The  simulations used state-of-art numerical techniques and subgrid physics including radiative cooling and photoheating \citep{Wiersma2009}, star formation law \citep{Schaye2008}, stellar evolution and enrichment  \citep{Wiersma2009b}, stellar feedback \citep{Dalla2012}, and black hole seeding and growth \citep{Matteo2005, Rosas2013}. These models were proven successful in reproducing many observed galaxy properties including the galaxy stellar mass function, galaxy sizes and the amplitude of the galaxy-central black hole mass relation, etc. \citep{Schaye2015}.    
The friends-of-friends(FOF) method \citep{Davis1985} was performed on the particle data to generate FOF groups by linking particles separated by 0.2 times the average particle separation. In each FOF group, the  SUBFIND \citep{Springel2001, Dolag2009} algorithm was applied to identify the self-bound particles as subhalos/substructures. M$_{200}$ is adopted to refer to the virial mass, which is the total mass within R$_{200}$ within which the average density is 200 times the critical density. Galaxies reside in the center of each substructure. The stellar mass is defined as the total mass of stellar particles within 30 pkpc radii of the centre of each subhalo. 
The luminosity and color are calculated using a stellar population synthesis model taking into account the SFR history and metallicity of each star particle \citep{Trayford2015}.  In the public EAGLE simulation catalog, it provides magnitudes in the five rest-frame  SDSS bands and  3 UKIRT bands. This sample contains absolute rest-frame magnitudes for all galaxies with $\rm M_{*} > 10^{8.5} \rm M_{\odot}$. No dust attenuation has been included.

\subsection{Environmental classification}
\label{sec:environment}

\begin{table*}
	\centering
	\label{tab:mass fraction table}
	\begin{tabular}{l cccc cccc}
    \toprule
    & \multicolumn{4}{c}{Volume fraction} &  \multicolumn{4}{c}{Mass fraction} \\
    \cmidrule(lr){2-5} \cmidrule(lr){6-9}
    Redshift    & Knot   & Filament  & Sheet & Void    & Knot   & Filament  & Sheet & Void \\
    \midrule
       0       & 0.010 & 0.168 & 0.416  & 0.406    & 0.279(0.268) & 0.393(0.398) & 0.239(0.244)  & 0.089(0.090)    \\
       1 & 0.010 & 0.138 & 0.397  & 0.455   & 0.154(0.147) & 0.355(0.358) & 0.329(0.332)  & 0.162(0.163)  \\
       3   & 0.007 & 0.083 & 0.331  & 0.579   & 0.042(0.042) & 0.209(0.209) & 0.392(0.392)  & 0.357(0.357)   \\
       6   & 0.001 & 0.024 & 0.187  & 0.788   & 0.004(0.004) & 0.057(0.057) & 0.272(0.272)  & 0.667(0.667)   \\
    \bottomrule
    \end{tabular}
	\caption{ 
    Volume fraction and mass fraction in each cosmic web component in the EAGLE simulations at $z \sim 0, 1, 3, 6$. Cols 1: redshifts; Cols 2, 3, 4, 5: volume fraction in knots, filaments, sheets and voids, respectively; Cols 6, 7, 8, 9: mass fraction in knots, filaments, sheets and voids, respectively (numbers within brackets are for baryonic mass fraction only). 
	}
\end{table*}

The cosmic web present itself in large-scale surveys, as well as in cosmological simulations. Many different approaches to classify the cosmic web elements have been developed in the literature.  
The tidal tensor and the velocity shear tensor are two of the most popular quantities to identify the cosmic web (see \citet{2018Libeskind} for a comparison of different classifications of the cosmic web). Here we follow \citet{Hahn2007} to use the tidal tensor to classify the four cosmic web elements.

A region is identified as void, sheet, filament or knot according to the number of dimensions that this particular patch collapses along. The tidal tensor $T_{ij}$ is given by the Hessian matrix of the gravitational potential $\phi$:
\begin{equation}
    T_{ij} = \frac{\partial ^2 \phi}{\partial r_i \partial r_j}.
	\label{eq:tidal tensor}
\end{equation}
The gravitational potential $\phi$ from the matter density field is obtained using the Poisson's equation:
\begin{equation}
    \nabla ^2 \phi = \delta, 
	\label{eq:poisson}
\end{equation}
where $\overline{\rho}$ denotes the mean mass density of the universe and  $\delta = \frac{\rho}{\overline{\rho}}-1$ denotes the density contrast. In practice, we split the simulation box into 256$^3$ cartesian cells and estimate the density and density contrast by assigning the particles to each cell using the cloud-in-cell method \citep{Sefusatti_2016}. We further smooth the discrete density field with a Gaussian filter of width $\rm R_s = 1.25{\rm Mpc}/h$.  

There are three eigenvalues of the tidal tensor T$_{i,j}$, $\lambda_1 \geq \lambda_2 \geq  \lambda_3$.  
\citet{Hahn2007} used the number of positive eigenvalues of T$_{ij}$ ($\lambda_{th}$= 0) to classify the four possible environments.
However, in reality, such criteria lead to an under-estimate of the filling factor of the voids. We follow \citet{Forero-Romero2009} to introduce an eigenvalue threshold $\lambda_{th}$, as a free parameter and define the cosmic web elements as following:

(\romannumeral1) Void: all eigenvalues $\lambda_i$ below the threshold $\lambda_{th}$ ($\lambda_{th} \geq \lambda_1 \geq \lambda_2 \geq  \lambda_3$).

(\romannumeral2) Sheets: two eigenvalues $\lambda_i$ below the threshold $\lambda_{th}$ ($\lambda_{1} \geq \lambda_{th} \geq \lambda_2 \geq  \lambda_3$).

(\romannumeral3) Filaments: one eigenvalues $\lambda_i$ below the threshold $\lambda_{th}$ ($\lambda_{1} \geq \lambda_{2} \geq \lambda_{th} \geq  \lambda_3$).

(\romannumeral4) Knots: all eigenvalues $\lambda_i$ above the threshold ($\lambda_{1} \geq \lambda_{2} \geq \lambda_{3} \geq  \lambda_{th}$).

Here we adopt  a fixed $\lambda_{th} = 0.25$ at $z = 0,1,3.01,5.97$. We test our results using different  $\lambda_{th}$ at different redshifts and find the results are in qualitative agreement with those with the fixed $\lambda_{th}$. This is consistent with the conclusions reached by \citet{2017Zhu}. In Fig.~\ref{fig:web}, we show the map of the overdensity and the corresponding cosmic web in a slice $0.265 {\rm Mpc}/h$ thick at $z = 0$. The cosmic structures are well captured by the tidal tensor method.

\begin{figure*}
\centering
\includegraphics[width=5.4in]{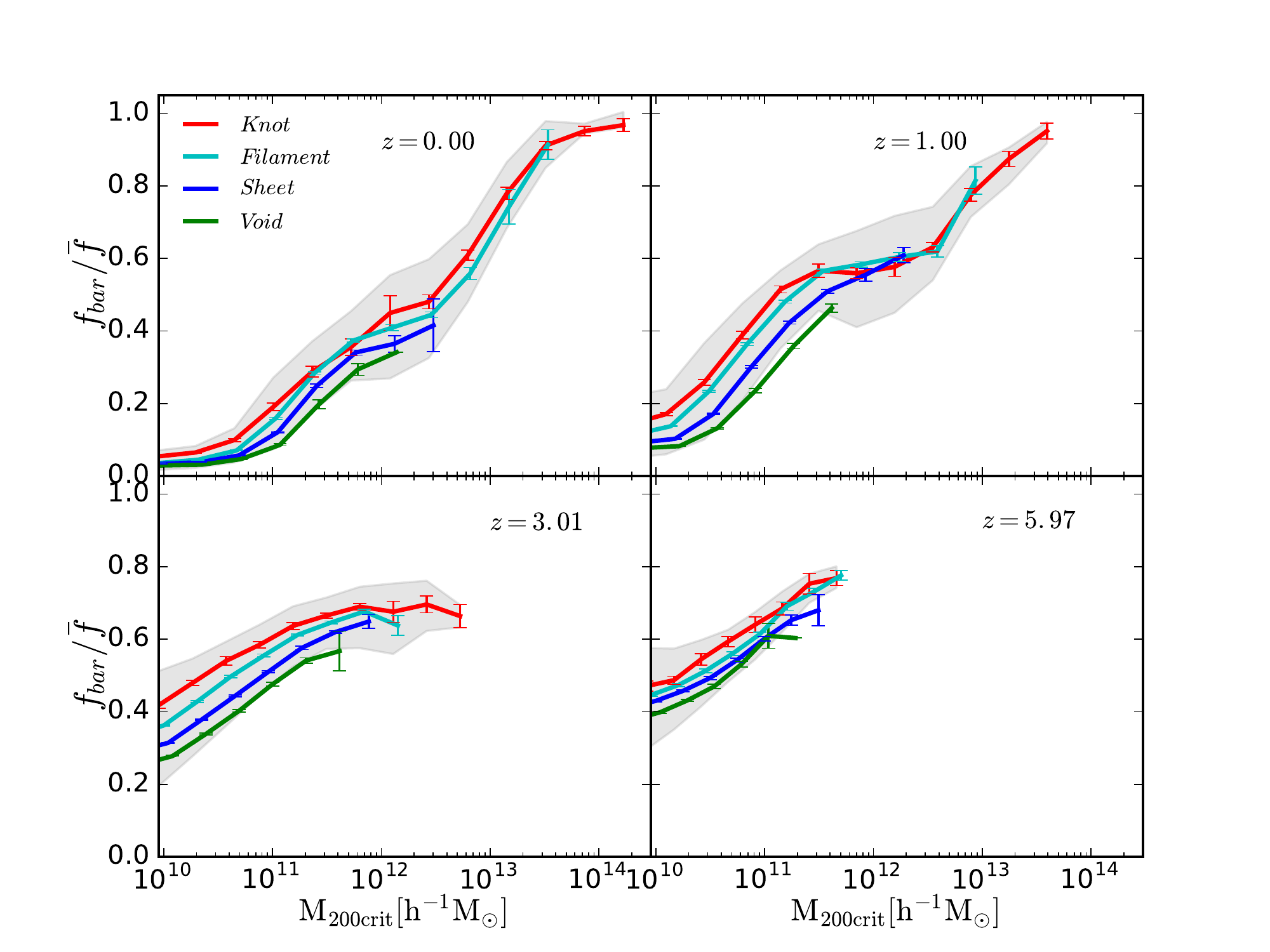}
\caption{Baryon fraction as a function of halo mass in different cosmic web environments. The shaded regions indicate the $16th$ -$84th$ percentile scatter in each corresponding halo mass bin. Red, cyan, blue and green curves donate the median values of the baryonic fractions in knots, filaments, sheets and voids, respectively.  Error bars present errors on the median values estimated based on 1000  bootstrap samples. Redshifts are indicated on the top right corner of each panel.}
\label{fig:baryon fraction}
\end{figure*}

\section{Baryonic properties in the cosmic web}
\label{sec:baryon}

\begin{figure*}
\centering
\includegraphics[width=5.3in]{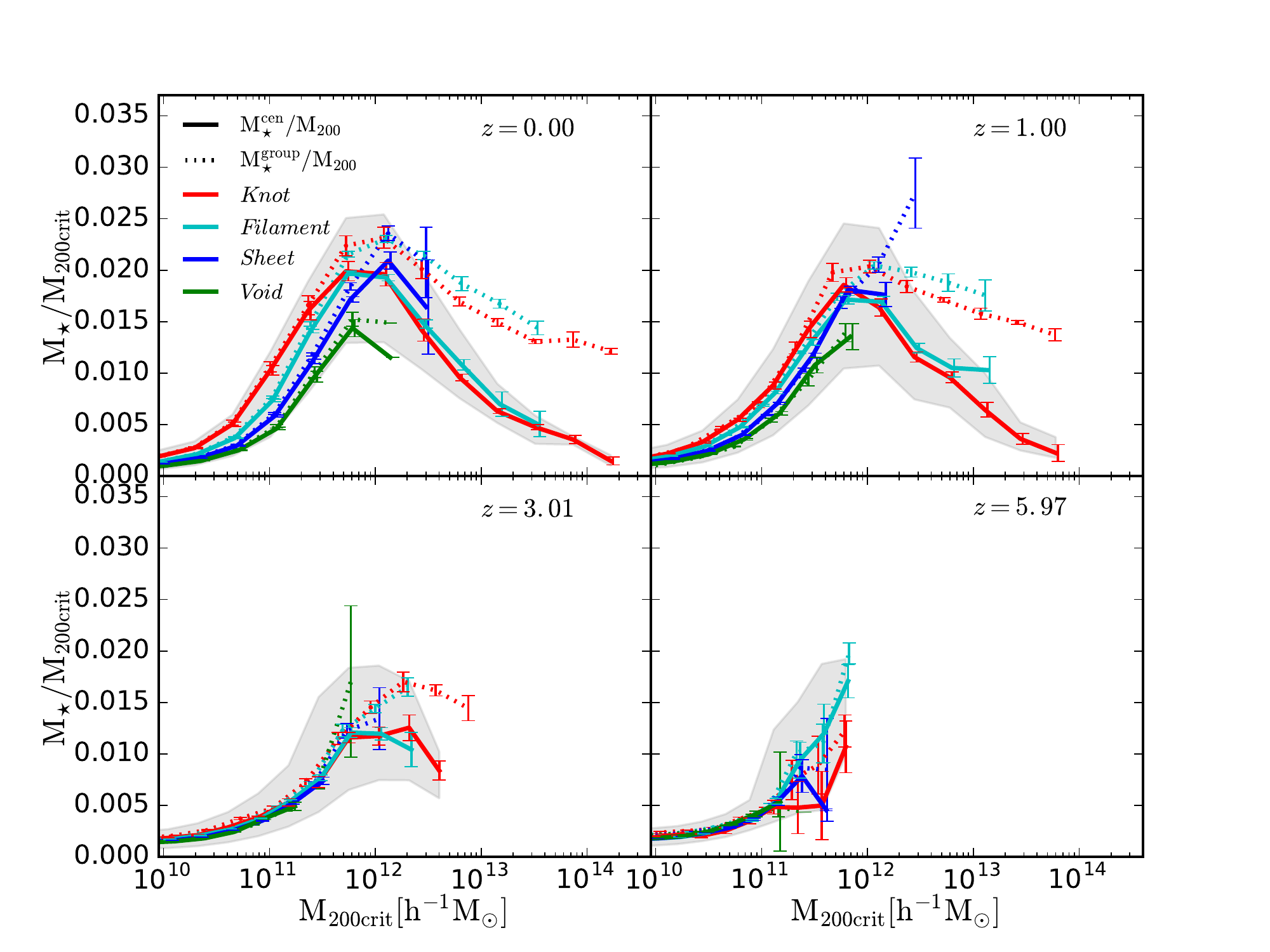}
\caption{Stellar mass-to-halo mass relation as a function of cosmic web environments. The solid curves account for stellar mass in central galaxies while the dashed curves account for the total stellar mass within the corresponding virial radius. The Shade regions indicate the $16th$ -$84th$ percentile scatter in each corresponding halo mass bin. Red, cyan, blue and green curves donate the median values of the ratios in knots, filaments, sheets and voids, respectively.  Errors are estimated based on 1000  bootstrap samples. Redshifts are indicated on the top right corner of each panel.}
\label{fig:stellar fraction}
\end{figure*}

\subsection{Total baryons}
Table 1 summarizes the volume fractions and mass fractions in different cosmic web environments. At z =0 the volume fractions are $0.407, 0.416, 0.168, 0.010$ among the void, sheets, filaments and knots, while the total mass fractions located in each structure element are $0.089, 0.239, 0.392, 0.279$. This is consistent with the results using Illustris simulation that most matter resides either in haloes or in filaments but very little in voids, though the voids contribute a significant fraction of the total volume \citep{Haider2016}. Both the volume fraction and mass fraction of voids increase with redshift, whilst they decrease in knots and filaments. 

For the baryon component, as expected, it is a good tracer of the total matter at large scales across cosmic time  \citep{Cen:1998hc,2001Dav,2015Eckert}. We find that the baryon fraction in knots and filaments decreases dramatically with redshifts. \citet{2019Cui} reached similar conclusions as ours, though their classification of the cosmic web adopted a different cell size and eigenvalue threshold, indicating our results are qualitatively robust against the method of classification of the cosmic web. In addition, \citet{2019Cui} demonstrated that the baryon fraction in different components of the web and its evolution depend only very weakly on the different physics implemented in the simulations. 

We further investigate the normalized baryon fraction ($f_{\rm bar}/\bar{f}$ = $({\rm M_{\rm bar}}/{\rm M_{200crit}})/( \Omega_b/ \Omega_0)$) as a function of halo mass in different cosmic web environments  (Fig.~\ref{fig:baryon fraction}), where $\bar{f}$ is the universal baryonic fraction 15.7\%, $f_{\rm bar}$ is the baryon fraction measured within R$_{200}$.
It shows that the baryonic fraction is an increasing function of the halo mass in all web environments. The strong halo mass dependence of the baryonic fraction presents itself over all the redshifts from 0 to 6. At z =0, only in massive clusters ($\sim 10^{14} h^{-1}\rm M_{\odot}$), the baryonic fraction does reach the universal value, while this fraction drops dramatically towards lower masses. At low masses, the overall baryonic fraction increases significantly with redshifts. 
For example, for haloes of  ${ \rm M_{200crit}} \sim 10^{10} h^{-1} \rm M_{\odot}$  the baryon fraction increases from $0.1$ at $z \sim 0$ to $0.4$ at $z \sim 5.97$.  In the MassiveBlack-II simulation, \citet{2015Khandai} found an even stronger redshift evolution of the baryon fraction.
\citet{2012Peirani} found a similar trend of the mean baryonic fraction in groups over cosmic time using hydrodynamical zoom-in simulations. They claimed that it is due to the lower accretion rate of dissipative gas onto the haloes compared to that of dark matter at low redshifts. Another reason for the high baryon fraction at high redshifts is that the potential well is deeper and it is thus harder to expel baryons out even for haloes of relatively low masses. On the other hand, for the massive systems (more massive than $10^{13} h^{-1} \rm M_{\odot}$) their baryonic fractions hardly change with redshifts. Haloes of such masses have potential wells deep enough to keep most of their baryons both at low and at high redshifts.

The baryonic fraction increases from voids, sheets, filaments towards knots. For haloes of $10^{11} h^{-1} \rm M_{\odot}$, the baryonic fraction in voids is lower than that in knots by 0.8 dex at z=0. It is much stronger dependence than that found by \cite{Metuki2015} who used simulations with different subgrid physics. This cosmic web dependence gets weaker towards lower masses and almost vanishes at $10^{10} h^{-1}  \rm M_{\odot}$. Fig. 2 also shows that in general the cosmic web dependence gets stronger with redshifts up to z =3, especially at low masses.  At the highest redshift, z$\sim$6, the cosmic web dependence gets weaker again. 

\subsection{Stellar mass-to-halo mass ratios}
In this section, we focus on the stellar mass and its relation to the cosmic web. The cosmic web dependences of the galaxy properties vs. stellar masses relations are presented in Sec.~\ref{sec:scaling}. 

We show the stellar mass-to-halo mass ratio for central galaxies ($\rm M_{\star, central}/M_{200crit}$)  at various redshifts in Fig.~\ref{fig:stellar fraction}. As presented by \citet{Schaye2015}, the trend of the full sample is consistent with that inferred by the subhalo abundance matching methods \citep[e.g.][]{Guo2011, 2013Moster} at z = 0. A similar analysis is applied to the total stellar mass fraction ($\rm M_{\star, group}/ \rm M_{200crit}$, dotted curves), where the total stellar mass referred to is the total stellar masses within R$_{200}$.  It peaks at $\sim 10^{12}h^{-1} \rm M_{\odot}$ and drops fast both towards the low mass and high mass ends. Different from the stellar-to-halo mass ratio for central galaxies, the slope at high masses is much flatter for the total stellar masses. At low masses, the central and total stellar mass to halo mass ratios are almost identical, indicating that satellite galaxies and intra-cluster lights contribute more to massive systems than to low mass systems. These trends are found at redshifts up to z $\sim$ 3. At z = 6 no high mass systems are formed due to the limited box size.  The amplitudes of the central and total stellar-to-halo mass ratios decrease towards higher redshifts, consistent with the finding from the subhalo abundance matching method  \citep{2013Moster} that the galaxy formation efficiency decreases slightly with increasing redshifts.

At z = 0 the central and total stellar mass-to-halo mass ratios decrease from knots to filaments to sheets and to voids for haloes below $\sim 10^{12} h^{-1} \rm M_{\odot}$. In haloes of $10^{11} h^{-1} \rm M_{\odot}$, where most of the dwarf galaxies reside, the ratios drop by a factor of $\sim$ 2. At high masses, it is at the opposite. In knots the ratios are the lowest while in voids the ratios are the highest. At the peak location ($\sim 10^{12} h^{-1} \rm M_{\odot}$), about the mass of the Milky Way's halo, the central and total stellar mass-to-halo mass ratios do not vary between different cosmic web environments.  Such environmental dependencies are also found at z=1, though somehow weaker than that at z=0. At even higher redshifts($z>3$), the cosmic web dependence vanishes at low masses, and no statistical results can be obtained at high masses due to the limited number of high mass systems. 

\begin{figure*}
\includegraphics[width=5.3in]{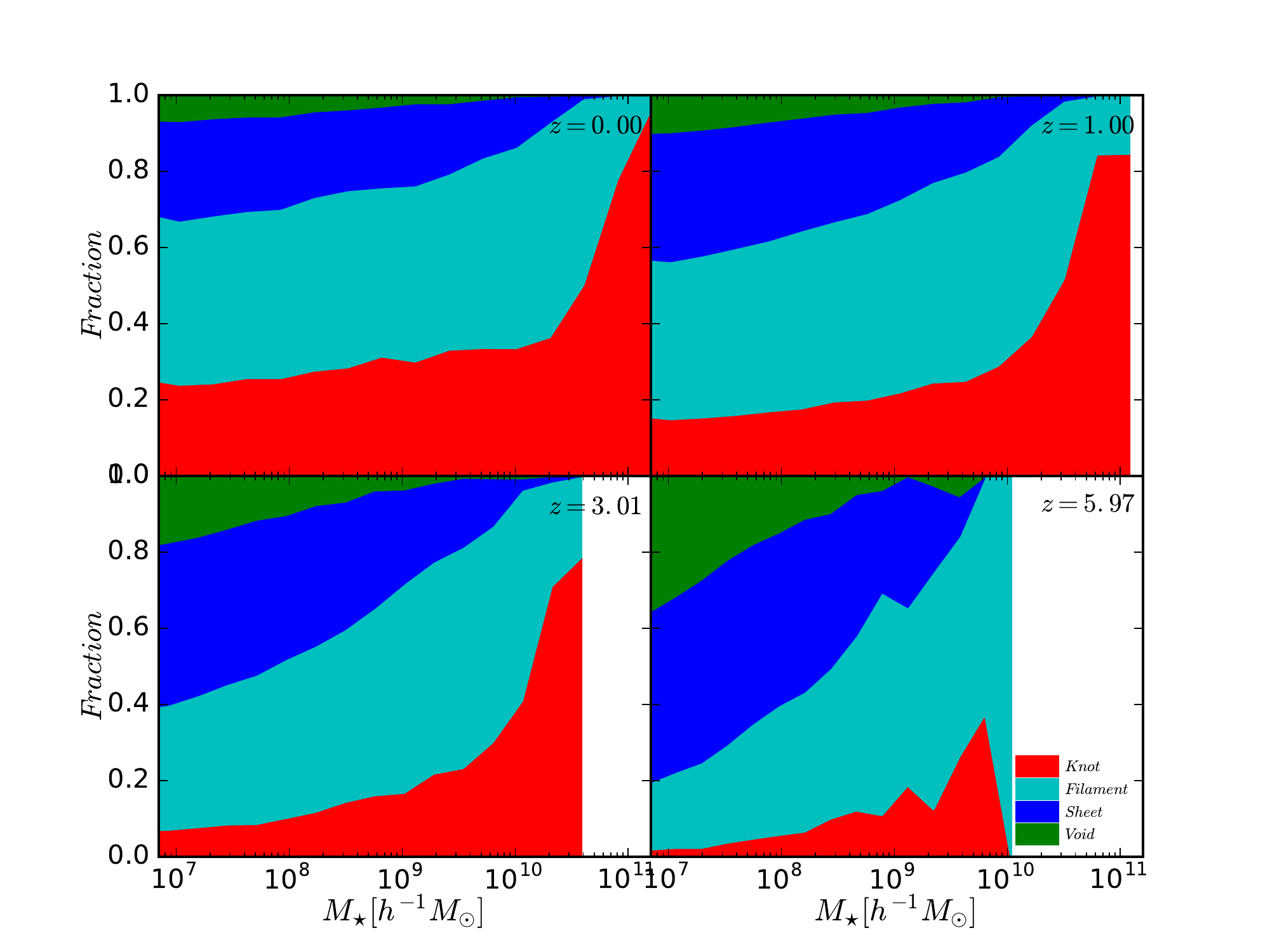}
\caption{Fractions of galaxies in different web environments as a function of stellar mass at $z \sim 0, 1, 3.01, 5.97$. Red, cyan, blue and green regions represent the fractions in knots, filaments, sheets and voids, respectively. Redshifts are indicated at the top right corner of each panel. }
\label{fig:fraction}
\end{figure*}

\begin{figure*}
\includegraphics[width=5.3in]{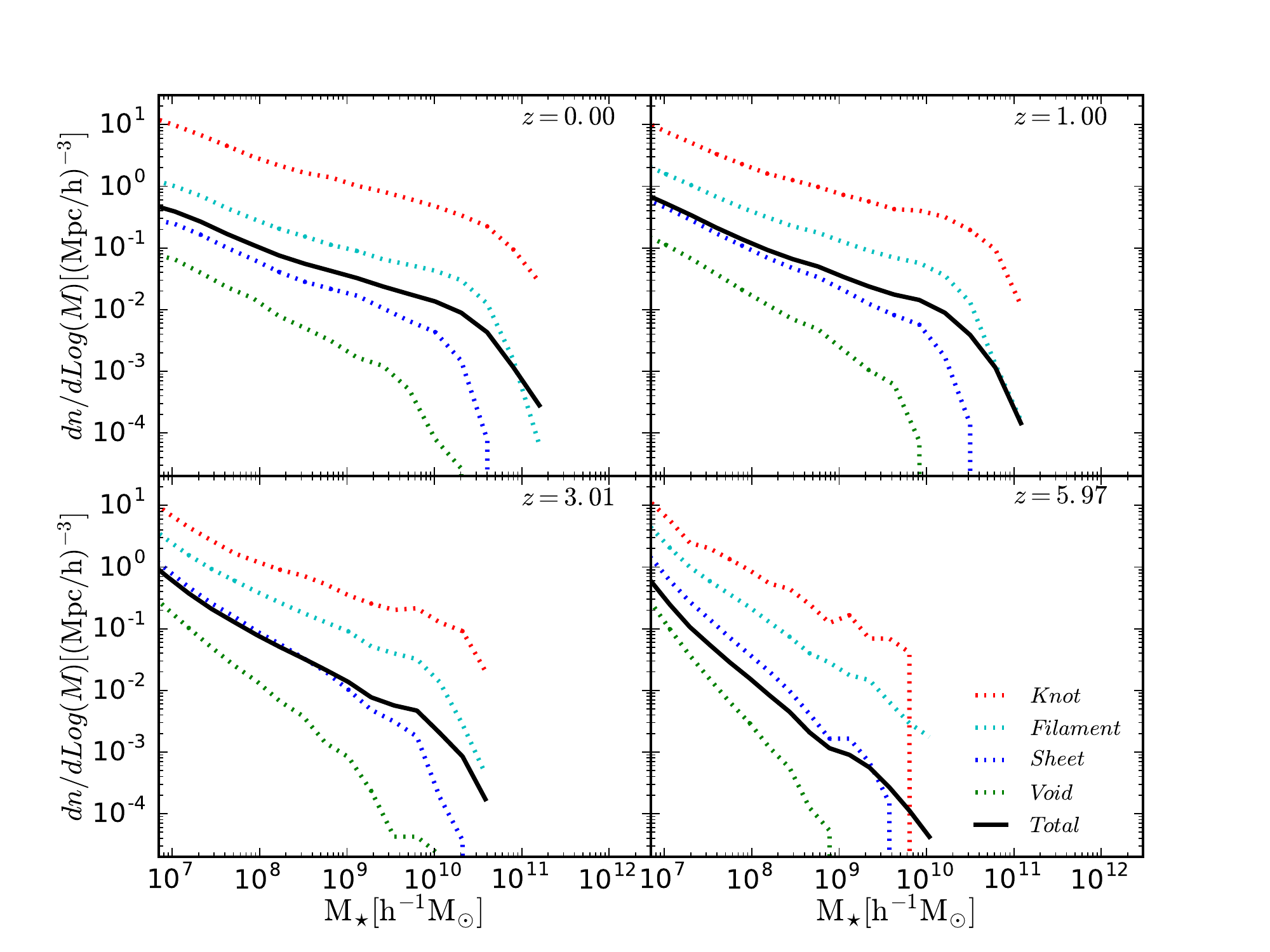}
\caption{The stellar mass functions in different cosmic web environments at different redshifts. Solid black curves are the overall stellar mass functions, while dashed ones are those for different environments with the same color coding as those in Fig.~\ref{fig:fraction}.  }
\label{fig:massfunc}
\end{figure*}

\subsection{Stellar mass functions}

Most stars are locked in galaxies. The fraction of galaxies as a function of stellar mass in different cosmic web environments are shown in Fig.~\ref{fig:fraction}. In a given stellar mass bin, the fraction of galaxies in any cosmic web environment is calculated by dividing the number of galaxies in the corresponding web environment by the total number of galaxies. In the top left panel, it shows that at z = 0 most of the massive galaxies reside in knots, consistent with previous results \citep{Metuki2015, Eardley2015}. For galaxies of the Milky Way mass and dwarf galaxies, most reside in filaments, though a comparable, yet slightly lower fraction reside in knots. This is also found by \citet{Eardley2015} using GAMA data and \citet{Metuki2015} using simulations.
Only a very small fraction of galaxies reside in voids at all masses considered here. The fractions in different cosmic web environments are similar at z = 1. At higher redshifts, more galaxies are found in voids and sheets than at lower redshifts, while less is found in knots and filaments, especially at low masses.

The dependence on the cosmic web is more significant when converting the total number of galaxies to the volume number density (the stellar mass functions) as shown in Fig. ~\ref{fig:massfunc}. At z =0, the number density drops by two orders of magnitude from knots to voids. Such strong dependences are also found by theoretical work \citep{Metuki2015} and observational work \citep{2015alpaslan}. Interestingly, we find the mass at the knee (corresponding to L$^*$) decreases by a factor of $\sim$ 8, with the value of $10^{10.8} h^{-1} {\rm M_{\odot}}$ in the knots and $10^{10.0} h^{-1}{\rm M_{\odot}}$ in voids. 
This change is dramatic given that the corresponding mass for the global stellar mass function barely changes since z $\sim $1 \citep[e.g.][]{2019Beare}.
As for the number fractions, the cosmic web environmental dependence of the stellar mass functions gets weaker at high redshifts. The evolution is stronger at low masses than that at high masses.

\begin{figure}
\centering
\includegraphics[width=3.5in]{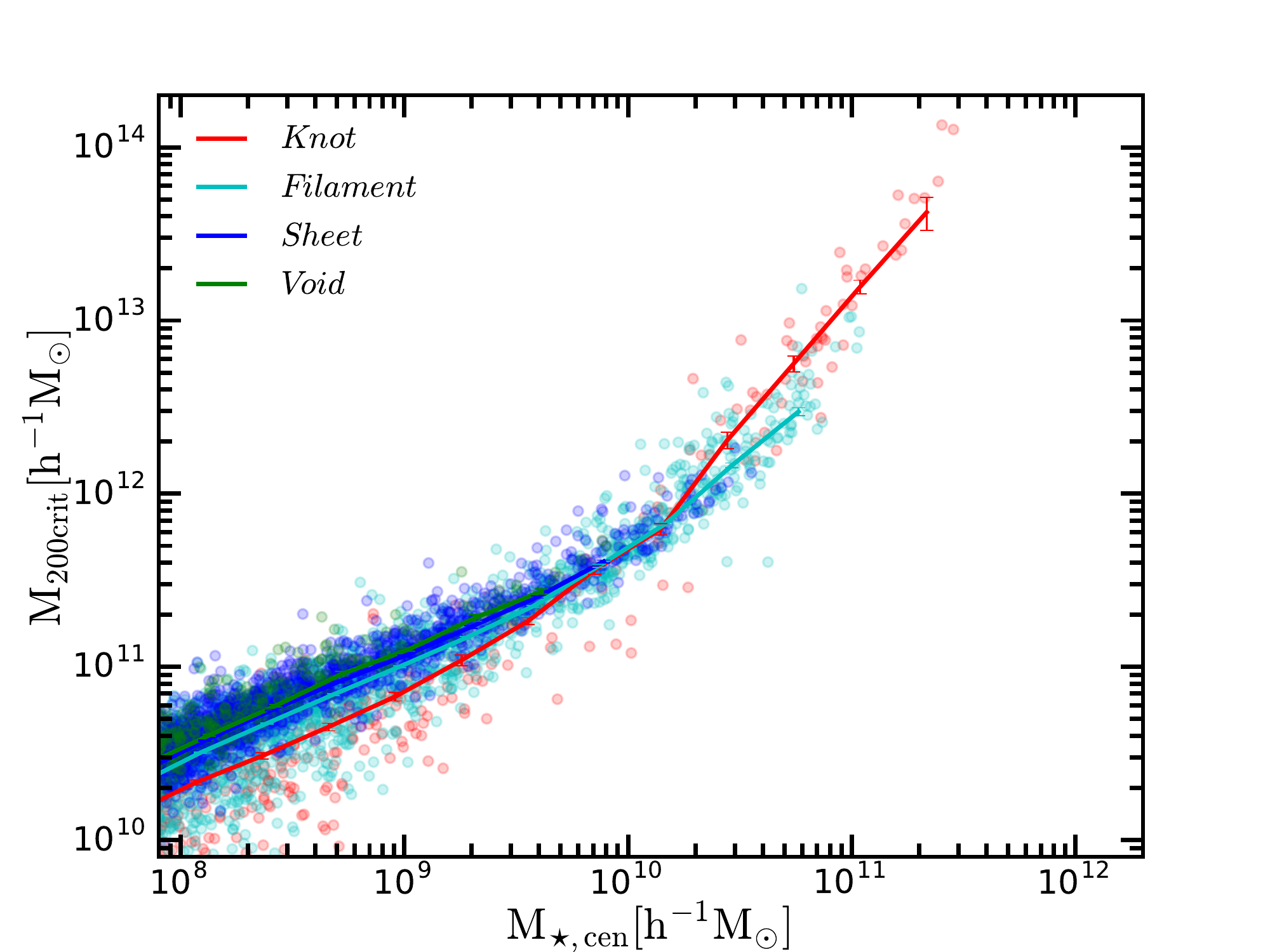}
\caption{The halo mass vs. stellar mass relation for central galaxies at z=0.  The Shade regions indicate the $16th$ -$84th$ percentile scatter in each corresponding stellar mass bin. Red, cyan, blue and green curves donate the median value of the ratios in knots, filaments, sheets and voids, respectively. Error bars are generated using the bootstrap method. }
\label{fig:stellar-halo mass relation}
\end{figure}

\begin{figure*}
\centering
\includegraphics[width=5in]{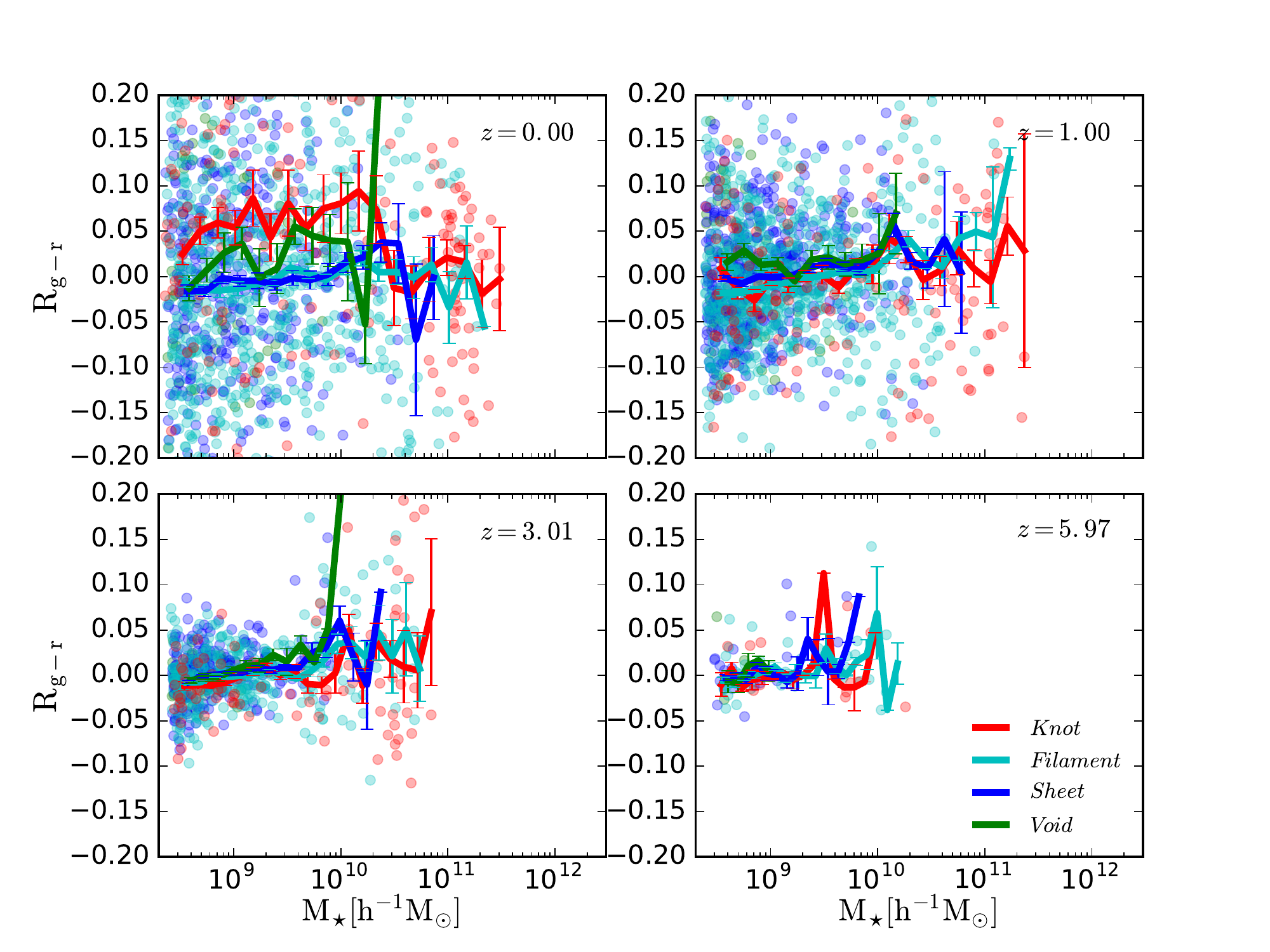}
\caption{The deviation of $g-r$ color from the expected values as a function of stellar mass for central galaxies in different environments. Red, cyan, blue and green curves donate the median value of the deviation in knots, filaments, sheets and voids, respectively. Scatters are for each galaxy with the same color coding as the curves. Error bars are generated using the bootstrap method. Redshifts are indicated at the top right corner of each panel.}
\label{fig:color evolution}
\end{figure*}

\begin{figure*}
\centering
\includegraphics[width=5in]{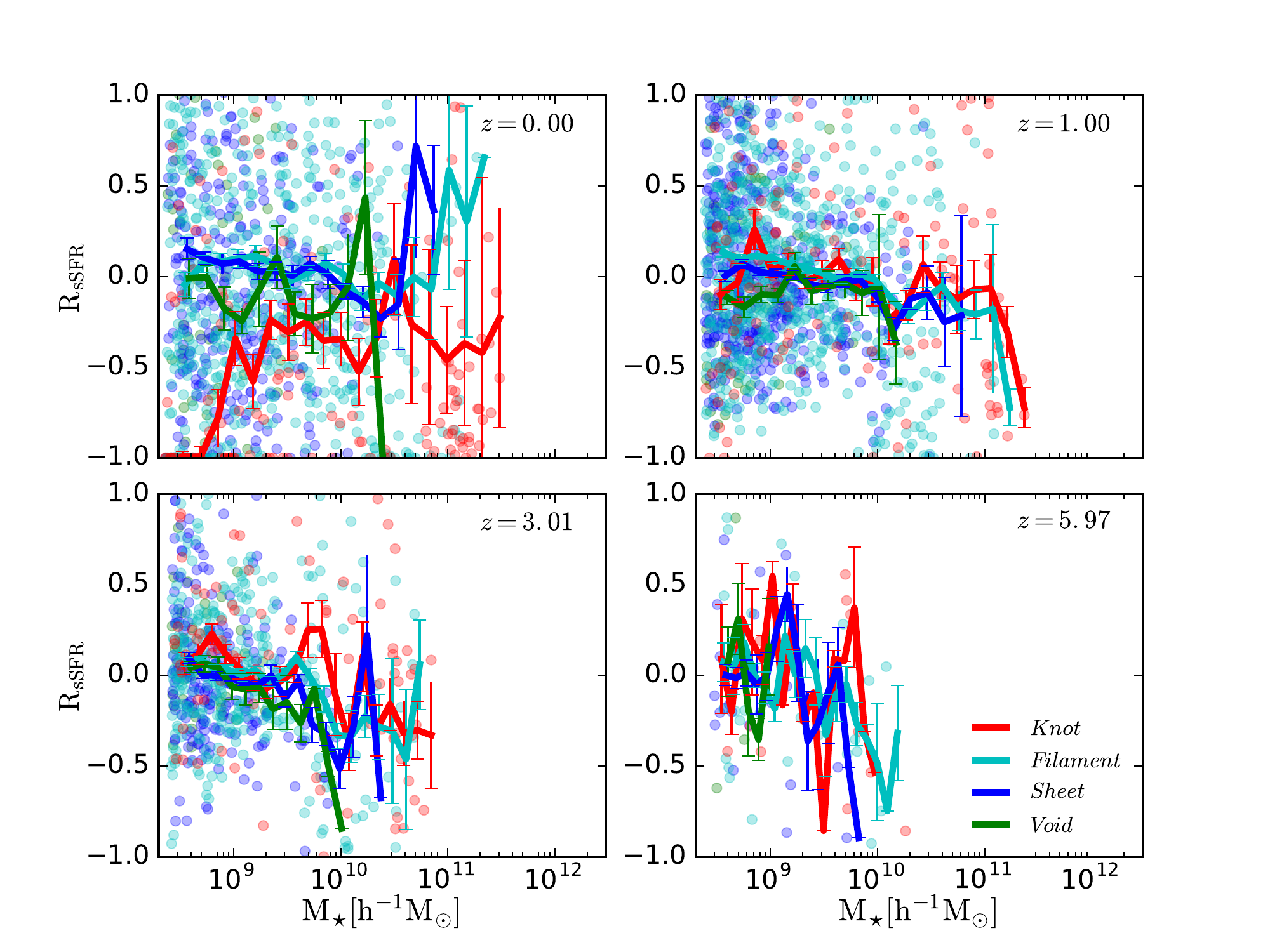}
\caption{The deviation of specific star formation rate from the expected values as a function of stellar mass for central galaxies in different environments. Red, cyan, blue and green curves donate the median value of the deviation in knots, filaments, sheets and voids, respectively. Scatters are for each galaxy with the same color coding as the curves. Error bars are generated using the bootstrap method. Redshifts are indicated at the top right corner of each panel.}
\label{fig:ssfr evolution}
\end{figure*}

\section{Scaling relations in the cosmic web}
\label{sec:scaling}

The observed scaling relations are very important in revealing the underlying physics of galaxy formation. In this section, we show how cosmic web environments affect the color vs. stellar mass relation, specific star formation rate (sSFR) vs. stellar mass relation and stellar/gas metallicity vs. stellar mass relation. Since the environmental dependence of satellite galaxies are very much different from that of central galaxies \citep[e.g.][]{2010Peng}, in this section we focus on {\it central} galaxies only. 

Previous work found that the cosmic web dependence and the halo mass dependence are highly degenerate \citep[e.g.][]{2016Brouwer}. Large scale web environments could shape the galaxy properties through the halo mass for galaxies properties depend strongly on halo mass \citep[e.g.][]{Metuki2015} and large scale structures affect halo masses significantly. We show in Fig.~\ref{fig:stellar-halo mass relation} that for a given stellar mass, the typical halo mass varies in different cosmic environments. For galaxies with stellar mass below $1.8\times$10$^{10}h^{-1}\rm M_{\odot}$, the host haloes are less massive in knots, while for those with stellar mass above $1.8\times$10$^{10}h^{-1}\rm M_{\odot}$, the host haloes are more massive in knots. This is consistent with what we found in Fig.~\ref{fig:stellar fraction}. In order to disentangle the connections between the cosmic web, dark matter haloes and galaxies, we measure the {\it pure} cosmic environmental effects by removing the dark halo effects. In practice, for any quantity of interest, $\rm X$, we calculate 
$\rm R_{\rm X}$ as a function of the cosmic web environment, where $\rm R_{\rm X}$ is defined as:

\begin{equation}
   \rm R_{\rm X} = \frac{\rm X}{\langle \rm X|\rm M_{200crit}\rangle} - 1,
	\label{eq:offset}
\end{equation}
For each $\rm M_{200crit}$ at a given redshift, we calculate the average value of $\rm X$, ${\langle \rm X|\rm M_{200crit}\rangle}$, in advance.  $\rm R_ {\rm X}$ quantifies the off-set of the property $\rm X$ that deviates from the expected values for given halo masses.

For any quantity $\rm X$ that depends directly only on halo mass, the mean of $\rm R_{\rm X}$ will equal 0 for any subsample, including one defined by stellar mass and environment, as below. By using $\rm R_{\rm X}$, we can test whether an apparent dependence of $\rm X$ on environment at a given stellar mass is entirely driven by the dependence of halo mass on environment at given stellar mass, or whether there is some remaining dependence on environment at fixed halo mass.

\subsection{Color vs. stellar mass}
\label{sec:color}

\begin{figure*}
\centering
\includegraphics[width=5.5in]{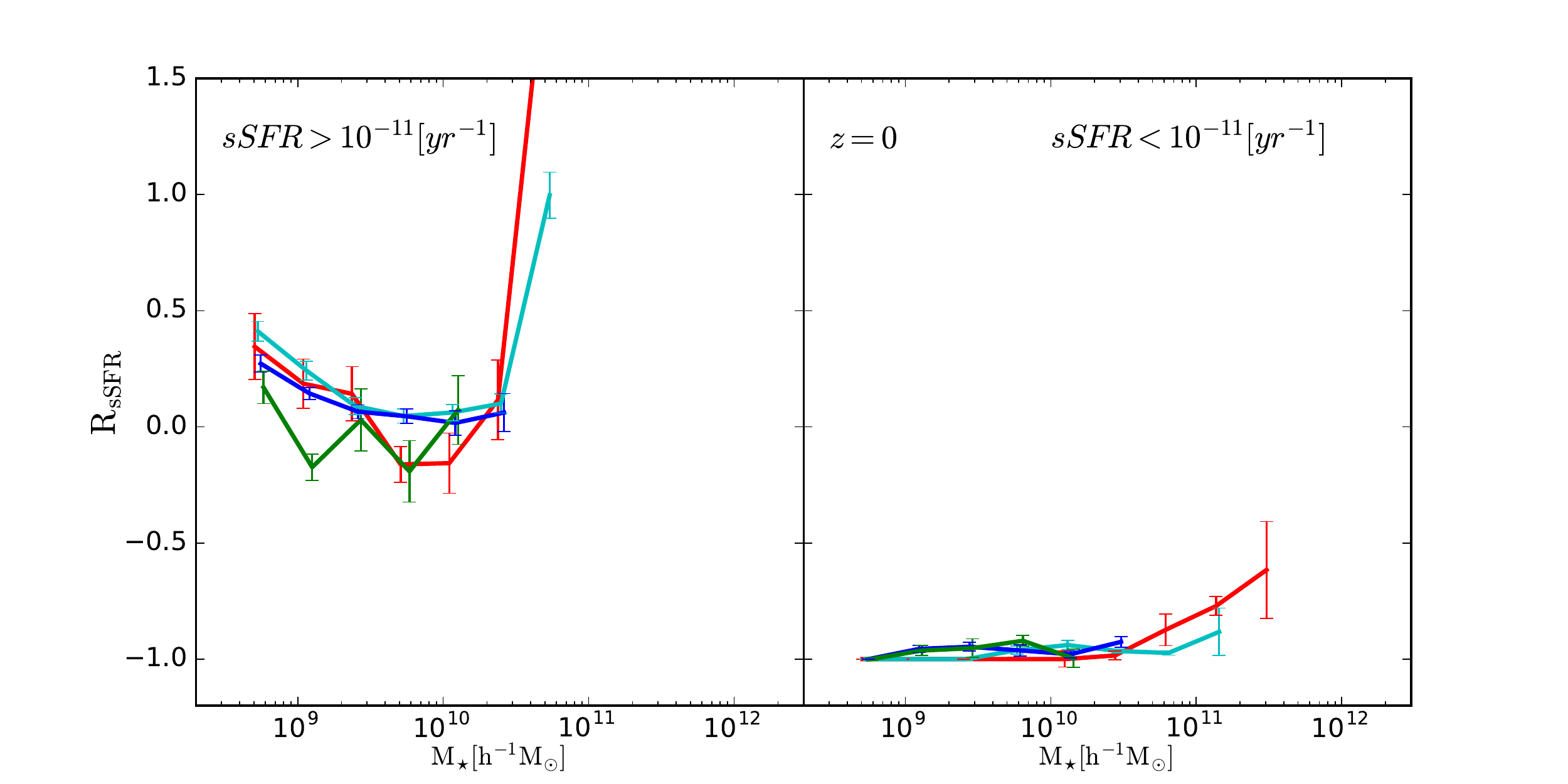}
\caption{The deviation of the sSFR from the expected values as a function of stellar mass in different environments for active (sSFR>$10^{-11}/yr$, left panel) and passive (sSFR<$10^{-11}/yr$, right panel) galaxies at z=0.  Red, cyan, blue and green curves donate the median value of the deviation in knots, filaments, sheets and voids, respectively. Error bars are generated using the bootstrap method.}
\label{fig:ssfr seprated}
\end{figure*}

Color is one of the most important observables for it is key to understand galaxies' star formation history. Previous studies found that galaxies on the color vs. stellar mass diagram can be grouped into three categories, blue cloud (active), red sequence (passive) and green valley (those in between).
Galaxies in low density environments (e.g. filaments, sheets and voids) tend to be bluer, while those in high density environments (knots) tend to be redder \citep{Poudel2017}. However, this could be caused by the fact that in low density regions galaxies and their host haloes are smaller, while smaller galaxies are in general bluer \citep{2011Cucciati, 2006Baldry}.

Fig.~\ref{fig:color evolution} shows the $g-r$  color vs. stellar mass relation as a function of the environments and stellar mass. Observed colors can be influenced by dust extinction and we thus use the intrinsic color. 
For the color is a dimensionless quantity and can be 0 in many cases, instead of using Eq. (3) we adopt a slightly different quantity to remove the halo dependence. For each galaxy, we calculate  $\rm R_{g-r}$  = (g-r) - ${\langle (g-r)|\rm M_{200crit}\rangle}$. 
At z=0, for galaxies less massive than 1.8$\times 10^{10} h^{-1} \rm M_{\odot}$, it shows a clear dependence on environments, especially for those in knots. Galaxies in knots are redder than those in voids by 0.08 $mag$ (1.8 $\sigma$). 
This is related to the fraction of active galaxies in different environments which we will discuss in more detail in the next subsection. At stellar masses above 1.8$\times 10^{10} h^{-1} \rm M_{\odot}$, there is no clear cosmic web environmental dependence. The difference between voids and sheets is weak across all the stellar mass ranges.

The dependence on environments vanishes for galaxies at high redshifts: z =1,3, and 6 at all masses. This is consistent with \cite{2017Darvish}  who found no cosmic web dependence of galaxy color in the COSMOS fields at z = 1.

\subsection{Specific star formation rate  vs. stellar mass}
\label{sec:ssfr}

Colors are influenced by the star formation history, especially the current star formation rate. Fig.~\ref{fig:ssfr evolution} shows the specific star formation rate (sSFR) as a function of stellar mass and cosmic environments. Similar to the color vs. stellar mass relation, at z=0, the cosmic environmental dependence is the strongest for galaxies with stellar mass less than $1.8\times 10^{10}h^{-1} \rm M_{\odot}$. The sSFR is lower by 50\% in knots than that in voids. At higher masses, given the large scatters, we can not find clear environmental dependence. This is consistent with the observational results by \citet{Poudel2017} who found central galaxies in low-density environments with higher sSFR compared to those in high-density environments at a fixed group mass using group catalogs \citep{2014Tempel} extracted from the SDSS DR10. The cosmic environmental dependence is much weaker at high redshifts, broadly consistent with the results of \citet{Scoville2013} and \citet{2016Darvish}. Different from the color vs. stellar mass relation, at z = 3 the sSFR is slightly higher in knots compared to that in other environments. 

Corresponding to the red sequence and blue cloud, galaxies can be separated into passive and active sub-populations using their specific star formation rates. Unlike galaxy colors, sSFR is not affected by metallicity and star formation history. Here we adopt log (sSFR[/yr])$ =$ -11 as the threshold: galaxies  with log (sSFR[/yr])$ <$ -11 are referred to as passive galaxies, while those  with log (sSFR[/yr])$ >$ -11 are taken to be active (star-forming) galaxies. The threshold value is fixed over cosmic time as its evolution is not very strong \citep{2019Matthee}.

\begin{figure}
\centering
\includegraphics[width=3.5in]{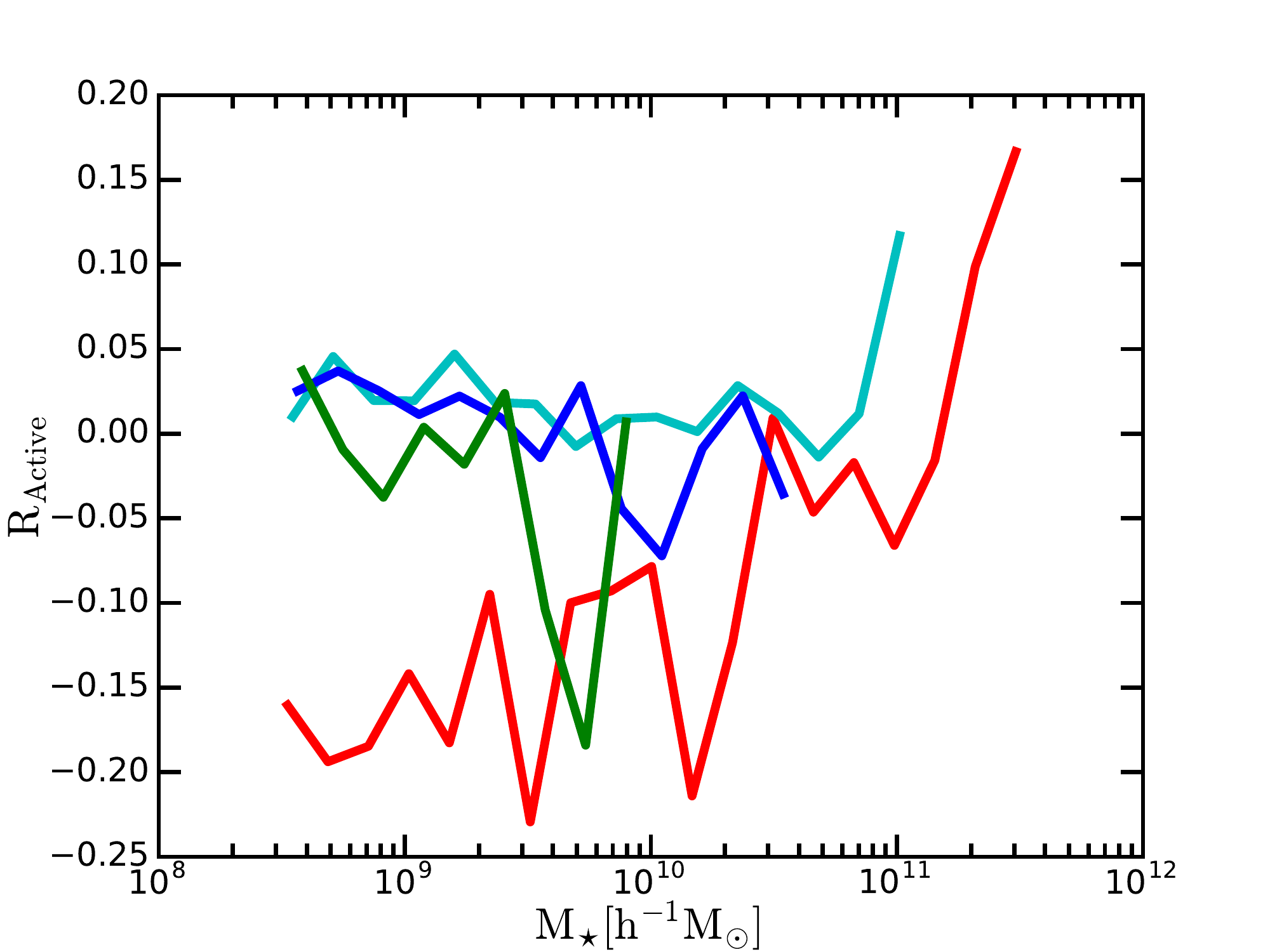}
\caption{The deviation of active fraction from the expected values as a function of stellar mass for central galaxies in different environments at z=0.  Red, cyan, blue and green curves are for the knots, filaments, sheets and voids, respectively.}
\label{fig:Blue fraction}
\end{figure}

\begin{figure*}
\centering
\includegraphics[width=5.3in]{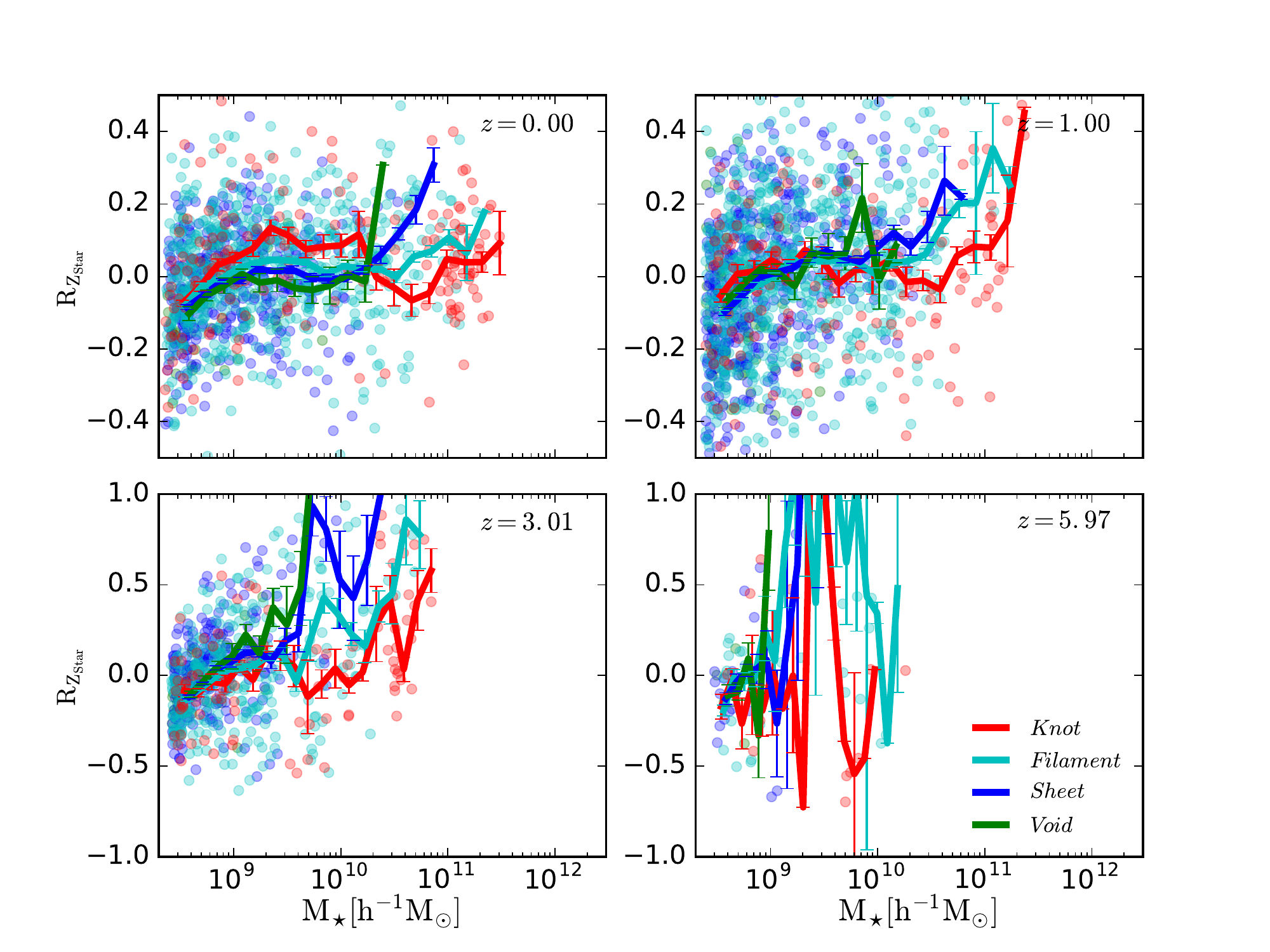}
\caption{The deviation of the stellar metallicity from the expected values as a function of stellar mass for central galaxies in different environments. Red, cyan, blue and green curves donate the median values of the deviation in knots, filaments, sheets and voids, respectively. Scatters are for each galaxy with the same color coding as the curves. Errors are generated using the bootstrap method. Redshifts are indicated at the top right corner of each panel.}
\label{fig:metallicity evolution}
\end{figure*}

The left panel in Fig.~\ref{fig:ssfr seprated} shows the environmental dependence of the sSFR for active galaxies at z=0.  For active galaxies, no environmental dependences present themselves except for those with masses $\sim 2.8\times 10^{9} h^{-1} \rm M_{\odot}$ where the sSFR of central galaxies is lower in voids than in other cosmic web environments. At higher masses, such dependence vanishes. For passive galaxies, as shown in the right panel, there is no environmental dependence over all the stellar mass ranges considered. 

The rather low sSFR in knots in the first panel of Fig.~\ref{fig:ssfr evolution} might be explained by their low fraction of active galaxies. The average active galaxy fraction increases from knots, filaments to sheets and voids, ranging from $0.58,0.81,0.87,0.88$, respectively. 
When taking into account the expected active galaxy fraction for any given halo mass and the number of haloes of the given mass in each cosmic web environment, the derived expected active galaxy fractions are $0.66, 0.78, 0.87, 0.91$ in knots, filaments, sheets, voids, respectively.
To make it more clear, we redo this analysis in each stellar mass bin and subtract the direct measurement of the active galaxy fraction from the correspondingly expected active galaxy fraction. 
The results are presented in Fig. ~\ref{fig:Blue fraction}. It shows that at masses below $ 1.8 \times 10^{10} h^{-1}\rm M_{\odot}$, there are less active galaxies in knots than in filaments, sheets and voids. It is the low fraction of active galaxies in knots that leads to the low sSFR in knots. It also explains the redder color in knots as shown in Fig.~\ref{fig:color evolution}. \citet{2011Sobral} have also argued that the environment is responsible for star-formation quenching in dense environments. In other words, denser environments increase the possibility of galaxies to become quenched. At high masses, on the other hand, there is instead not much difference in the active fraction in different environments. As a consequence, the cosmic web dependences of the color and sSFR also vanish. 

At high redshifts, most galaxies are star-forming and there is almost no environmental dependence of the active fraction: at $z>1$, the active fraction approaches 1 in all cosmic web environments. The environmental dependence of the sSFR thus vanishes.

\subsection{Metallicity  vs. Stellar mass}
\label{sec:metallicity}

\begin{figure*}
\centering
\includegraphics[width=5.3in]{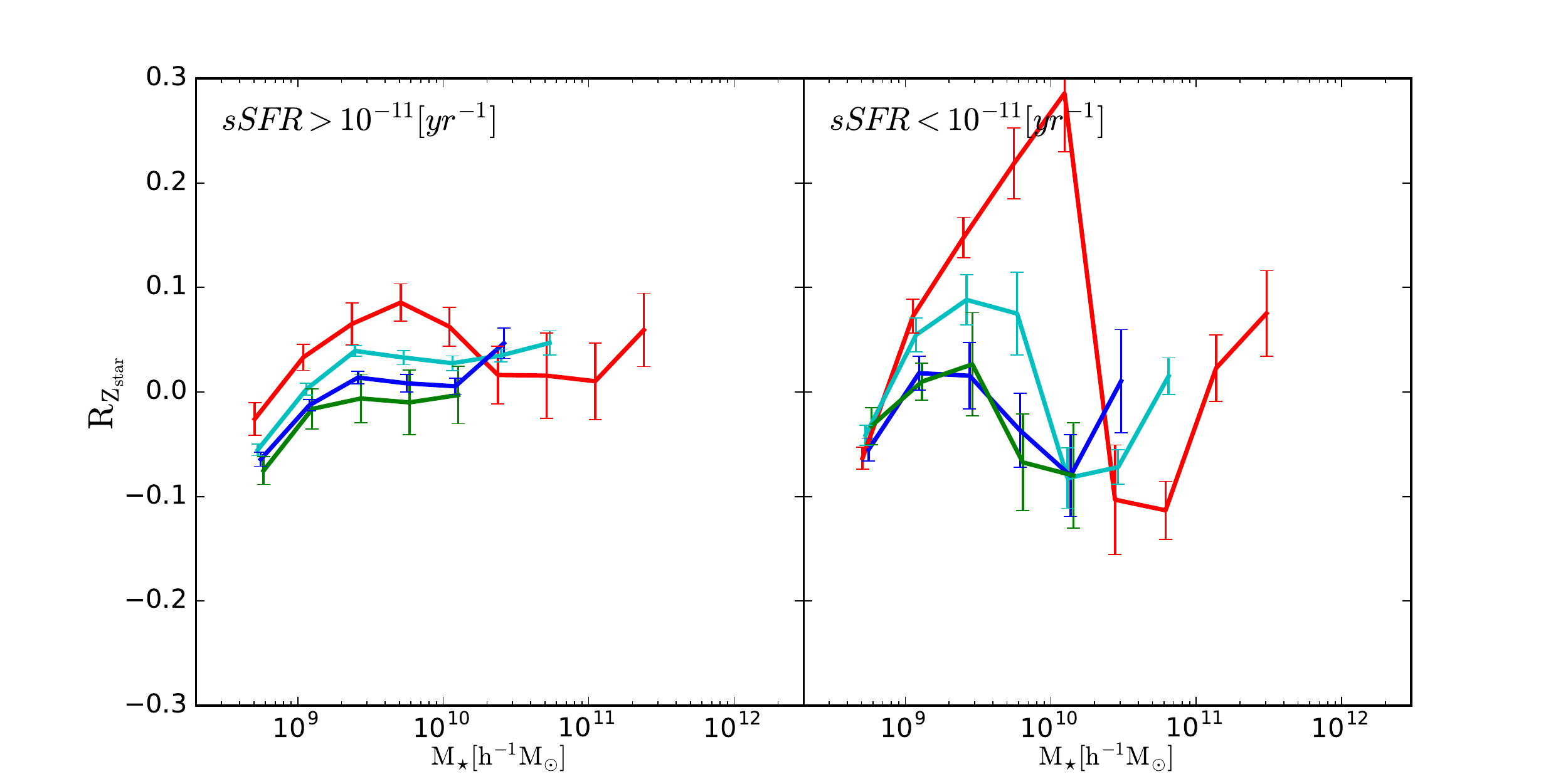}
\caption{The deviation of the stellar metallicity from the expected values as a function of stellar mass in different environments for active (left panel) and passive (right panel) galaxies. Red, cyan, blue and green curves donate the median values of the deviation in knots, filaments, sheets and voids, respectively. Errors are generated using the bootstrap method. }
\label{fig:metallicity seprated}
\end{figure*}

\begin{figure*}
\centering
\includegraphics[width=5.3in]{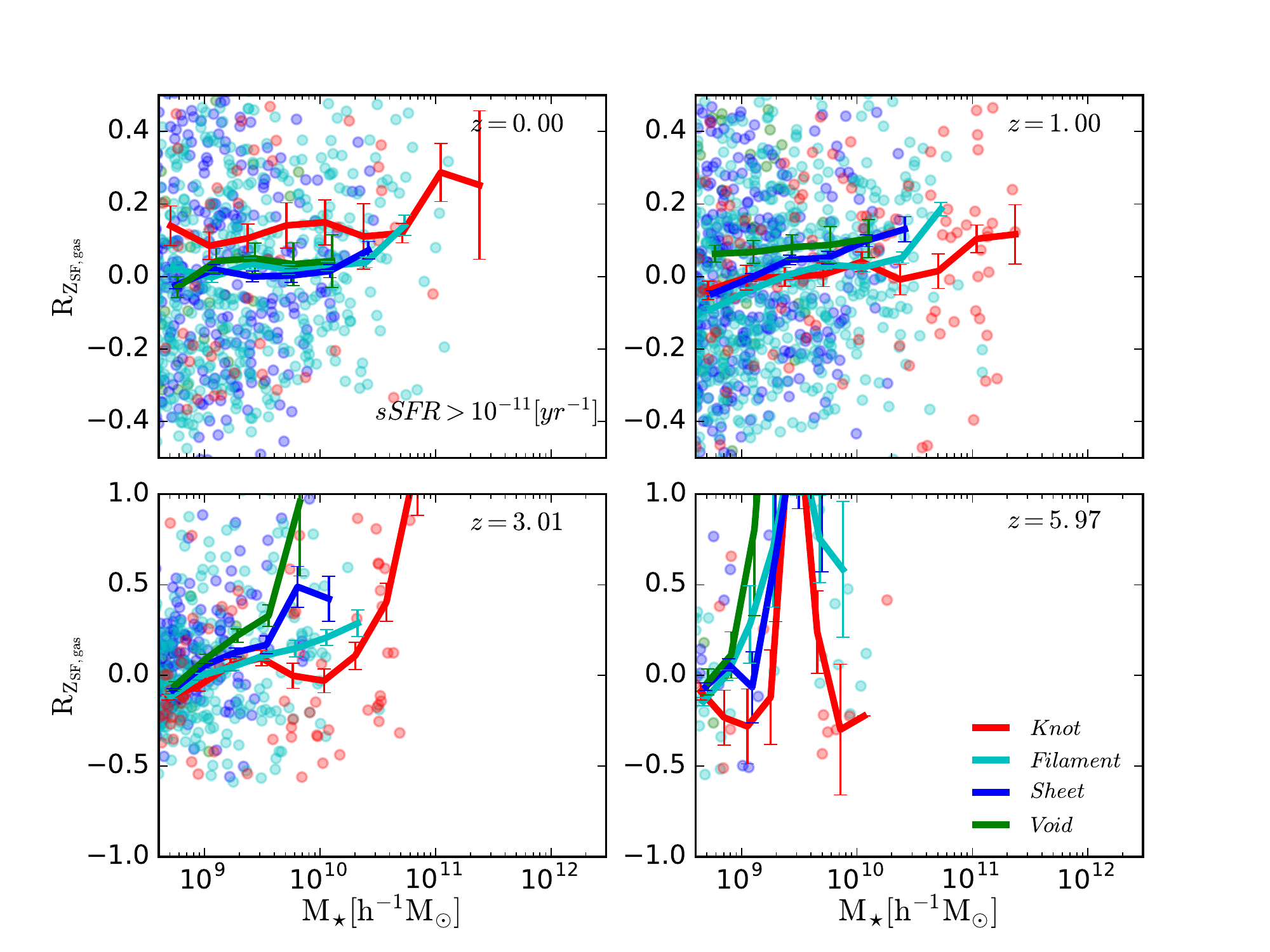}
\caption{Similar to Fig.~\ref{fig:metallicity evolution} but for gas metallicity of active galaxies.}
\label{fig:metal SFgas}
\end{figure*}

Metal enrichment is one of the most important processes in galaxy evolution which involves the gas cooling, star formation, supernova feedback, etc. \citet{Schaye2015} found the gas and stellar metallicity vs. stellar mass relations in EAGLE are in broad agreement with observations for galaxies more massive than  $10^{9} \rm M_{\odot}$,  though at the low mass end the relations are not as steep as the observed ones.  

Fig.~\ref{fig:metallicity evolution} shows the stellar metallicity vs. stellar mass relation in the different cosmic web environment.  At low masses (${\rm M_{\star}} \sim 1.8 \times 10^{10} h^{-1}\rm M_{\odot}$) the metallicity is higher in knots  than in other environments. This is related to the relation between stellar mass, metallicity and star formation rate, as discovered in previous works \citep[e.g.][]{2008Ellison,2010Mannucci,2012Yates,2015DeRossi}. Specifically, \citet{2017DeRossi} found that in EAGLE simulations the metallicity of low-mass systems decreases with SFR  for a given stellar mass. The high metallicity in knots is thus consistent with their low star formation rates as shown in Fig.~\ref{fig:ssfr evolution}. Such environmental dependence is much weaker at high redshifts.

Different from the relation with sSFR and color, the stellar metallicity vs. stellar mass relation shows a strong dependence on cosmic environments at high masses. The metallicity increases from knots, filaments, sheets towards voids. This is consistent with the increasing central stellar mass-to-halo mass ratios along knots, filaments, sheets and voids, i.e. more metals are generated if more stars are formed. Such dependence persists up to z=3. At even higher redshifts, there are not enough samples to make solid conclusions. 

As for the sSFR, we split the galaxies into active and passive subsamples at z = 0. The transition mass at ${\rm M_{\star}} \sim 1.8 \times 10^{10} h^{-1}\rm M_{\odot}$ presents itself both for the active and passive galaxies, below which the metallicity is higher in knots while above which the metallicity is lower in knots. The environmental dependence is stronger for passive galaxies than for active galaxies. 

The cosmic web dependences of the gas metallicity as a function of stellar mass and redshift are presented in Fig.~\ref{fig:metal SFgas}. Since there is very little gas in passive galaxies,  here we focus on active galaxies. At z=0, at stellar masses below $1.8 \times 10^{10} h^{-1}\rm M_{\odot}$ the gas metallicity is higher in knots, while at high masses the difference between different web environments disappears. Different from other scaling relations, this cosmic web dependence of gas metallicity gets stronger at higher redshifts for massive galaxies.  The gas metallicity is the lowest in knots and gets higher along filaments, sheets and voids. 

 \subsection{Combined effects of cosmic web and dark matter haloes}
\label{sec:webandhalo}

\begin{figure*}
\centering
\includegraphics[width=6.3in]{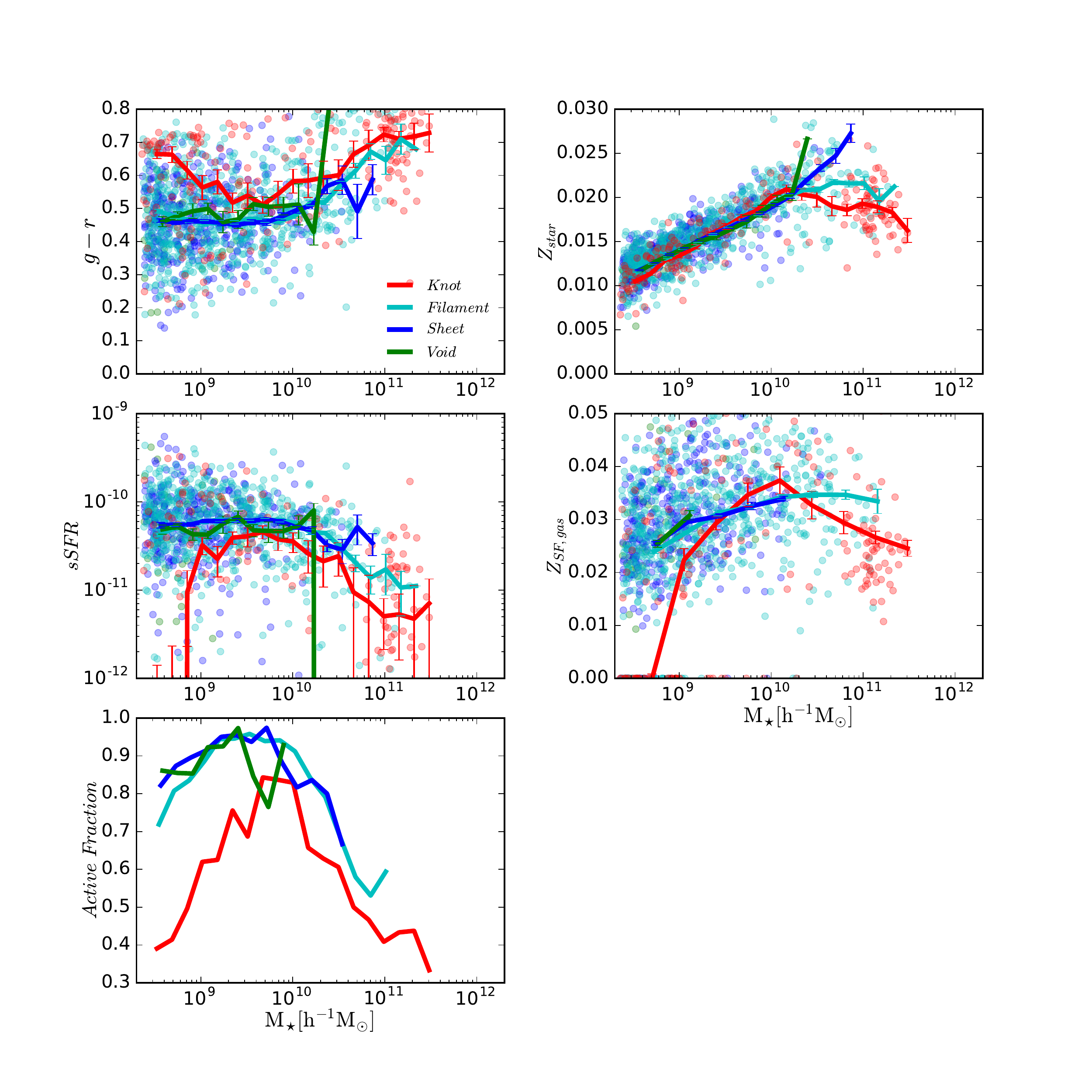}
\caption{The $g-r$ color, sSFR, active fraction, stellar metallicity and star forming gas metallicity as a function of stellar mass for central galaxies in different environments. Red, cyan, blue and green curves donate the median values in knots, filaments, sheets and voids, respectively. Each dot represent an individual galaxy with the same color coding as the curves. Errors are generated using the bootstrap method. }
\label{fig:galaxy properties}
\end{figure*}

We remove the halo effect from the cosmic web dependence in previous sections. However, it is difficult to obtain halo mass observationally. Here we present the {\it apparent} cosmic web dependence (including halo effects, hereafter we refer to it as combined dependence) on the scaling relations.  Since the cosmic environmental effect is weak at high redshifts for most of the scaling relations, we focus on results at z=0. 

The top left panel of Fig.~\ref{fig:galaxy properties} shows the g-r color vs. stellar mass relations for galaxies in different environments. Galaxies are redder in knots compared to other environments at all masses. This is different from the results shown in Fig.~\ref{fig:color evolution} that when removing halo effects at stellar mass above $1.8 \times 10^{10} h^{-1}\rm M_{\odot}$ galaxies in knots have similar colors as those in filaments, sheets and voids.  Such dependences are also found in the sSFR vs. stellar mass relations (middle left panel). These can be explained by the fact that there are more massive haloes in high density regions and galaxies in massive haloes are usually redder/with lower sSFR. The active fraction increases with stellar mass up to $\sim 3 \times 10^9 h^{-1}\rm M_{\odot}$ and decreases towards higher masses. This variations with stellar mass are similar to each other between voids, sheets and filaments. For those in knots, the amplitude is lower, the turn-over mass is higher, and at low masses the slope is steeper. 

For a given stellar mass, the halo mass is lower in knots for those with stellar mass below $1.8 \times 10^{10} h^{-1}\rm M_{\odot}$ (Fig.~\ref{fig:stellar-halo mass relation}). In low mass haloes the feedback is usually more effective and it is thus easier for new formed heavy elements to escape. This compensate with the cosmic web dependence of stellar metallicity as shown in Fig.~\ref{fig:metallicity evolution}, resulting in the absence of variance between different environments as shown in the right top panel of Fig.~\ref{fig:galaxy properties}. At high masses,  the stellar metallicity decreases with environmental densities, similar to those without halo effects (Fig.~\ref{fig:metallicity evolution}). For the gas metallically vs. stellar mass relation (right bottom panel  of Fig.~\ref{fig:galaxy properties}), there is almost no difference between voids, sheets and filaments. On the other hand, the scaling relation in knots is very different. It increases rapidly with stellar masses below $1.8 \times 10^{10} h^{-1}\rm M_{\odot}$  and also decreases towards high masses. The absolute values are lower in knots both at low and high masses compared to other environments, but is higher at the turn over mass.

In summary, for the color vs. stellar mass relation and sSFR vs. stellar mass relation, the combined dependence follows the pure cosmic web dependence at low masses quantitatively, while at high masses, the combined dependence is stronger. For the stellar metallicity vs. stellar mass, the combined dependence mimic the pure cosmic web dependence at high masses, but vanish at low masses. The gas metallically vs. stellar mass relation behaves significantly different in knots compared to other environments, and the combined dependence deviate from the pure cosmic web dependence at all masses.

\section{Conclusions}
\label{sec:conclusions}

In this paper, we investigate the dependence of galaxy properties on different cosmic web environments and their evolution, using the EAGLE cosmological hydrodynamical simulations. We split the simulation box into  $256^3$ cells and generate the web elements adopting the web classification method of \citet{Hahn2007}. Here we summarize our results as follows. 

We find the baryon fraction increases with halo mass in all environments, and the fraction is higher in denser regions, i.e. increasing along the sequence voids, sheets, filaments and knots. This environmental dependence persists up to redshift 6. The cosmic web dependence becomes slightly stronger at higher redshifts up to $z\sim3$ and then becomes weaker towards even higher redshifts. The central and total stellar mass-to-halo mass ratios both peak at halo masses $\sim 10^{12} h^{-1} \rm M_{\odot}$. At low masses, more stars are formed in knots than in other web environments, while at high masses, less stars are formed in knots.  The cosmic web dependence of the galaxy stellar mass functions is very strong at all redshifts, with the amplitude decreasing along the sequence knots, filaments, sheets and voids. Interestingly, though the average characteristic stellar mass corresponding to L$^*$ does not evolve much since z=1, it changes by an order of magnitude going from knots to voids at z =0.  

We remove the halo mass dependence and investigate the relation between the cosmic web and various scaling relations for central galaxies.
We find a characteristic stellar mass of $1.8\times 10^{10} h^{-1}\rm M_{\odot}$,  below and above which the cosmic web dependence behaves oppositely. 
Galaxies with stellar mass below the characteristic mass are redder, with lower active fraction, lower sSFR, and higher stellar metallicity in knots than in voids, while at stellar masses above the characteristic mass the dependences on the cosmic web either reverse or vanish. At low masses, the relatively strong web dependences of the color vs. stellar mass relation and of the sSFR vs. stellar mass relation can be attributed to the cosmic web dependence of the active galaxy fraction, i.e. the active fraction is higher in voids than in knots. For active galaxies, the stellar metallicity is higher in knots compared to other web environments for those with stellar mass below the characteristic mass.

The cosmic web dependences are weaker at higher redshifts for almost all the galaxy properties and scaling relations we explored, including the central (total) stellar-to-halo mass ratio, color vs. stellar mass relation, sSFR vs. stellar mass relation and stellar metallicity vs. stellar mass relation. But this is not the case for the gas metallicity vs. stellar mass relation. For galaxies above the characteristic stellar mass, this cosmic web dependence gets even stronger at high redshifts, decreasing along the sequence voids, sheets, filaments and knots.

The combined halo $+$ cosmic web dependence follow the cosmic web dependence at low masses for the color/sSFR vs. stellar mass relations, but is stronger at high masses. It reverses for the stellar metallicity at high masses, the combined dependence mimic the cosmic web dependence, while at low masses, the combined dependence almost vanish. One can not use the combined dependence of  gas metallicity relation vs. stellar mass relation to explore the pure cosmic web dependence at all for they behave different at all masses.

\section*{DATA AVAILABILITY}

The data presented in this article are available at the EAGLE simulations public database (\url{http://icc.dur.ac.uk/Eagle/database.php}).

\section*{Acknowledgements}

We thank Yingjie Jing and Marius Cautun for helpful discussion. This work is supported by the National Key $ R \&  D$ Program of China (Nos 2018YFA0404503, 2016YFA0400703 and 2016YFA0400702) and the National Natural Science Foundation of China(NSFC)(11573033, 11622325, 11133003, 11425312, 11733010, 11633004). L.G. also acknowledges the National Key Program for Science and Technology Research Development (2017YFB0203300). C.G.L was supported by the Science and Technology Facilities Council [ST/P000541/1]. X.C. also acknowledges the NSFC-ISF joint research program No. 11761141012, the CAS Frontier Science Key Project QYZDJ-SSW-SLH017, Chinese Academy of Sciences (CAS) Strategic Priority Research Program XDA15020200, and the MoST 2016YFE0100300, 2018YFE0120800. Q.G., L.G. and C.G.L acknowledge support from the Royal Society Newton Advanced Fellowships. This equipment was funded by BIS National E-infrastructure capital grant ST/K00042X/1, STFC capital grant ST/H008519/1, and STFC DiRAC Operations grant ST/K003267/1 and Durham University. DiRAC is part of the National E-Infrastructure.




\bibliographystyle{mnras}
\bibliography{cosmic} 




\bsp	
\label{lastpage}
\end{document}